\documentstyle{book}
 
\catcode`\@=11
 
\def\section{\@startsection {section}{1}{\z@}{3ex plus 1ex minus
.2ex}{1.2ex plus .2ex}{\normalsize\bf}}
\def\subsection{\@startsection{subsection}{2}{\z@}{3ex plus 1ex minus
.2ex}{1.2ex plus .2ex}{\normalsize\bf}}
 
\def\tableofcontents{\@restonecolfalse\if@twocolumn\@restonecoltrue
\onecolumn\fi\PROCcont\@starttoc{con}\if@restonecol\twocolumn\fi}
 
\def\l@section{\@dottedtocline{1}{0em}{.66em}}
 
\def\thebibliography#1{\section*{{\tenrm REFERENCES}\@mkboth
 {BIBLIOGRAPHY}{BIBLIOGRAPHY}}\list
 {[\arabic{enumi}]}{\settowidth\labelwidth{[#1]}\leftmargin\labelwidth
 \advance\leftmargin\labelsep\usecounter{enumi}}
 \def\newblock{\hskip .11em plus .33em minus -.07em}
 \sloppy\clubpenalty4000\widowpenalty4000
 \sfcode`\.=1000\relax}

\@addtoreset{footnote}{section}
 
\def\ps@myheadings{\let\@mkboth\@gobbletwo
\def\@oddhead{\hbox{}\hfil{\footnotesize\rm\rightmark}\hfil
\normalsize\rm\thepage}\def\@oddfoot{}\def\@evenhead{\normalsize\rm
\thepage\hfil{\footnotesize\rm\leftmark}\hbox{}\hfil
}\def\@evenfoot{}\def\sectionmark##1{}\def\subsectionmark##1{}}
 
\font\rompF=cmmi6
\def\ps@romp{\let\@mkboth\@gobbletwo
  \def\@oddhead{\small\rm Vol.\ \Volume\ (\Year)\hfil
  {\rompF REP\kern.5ptO\kern-.6pt
RTS\hspace{1.2ex}O\kern-.3ptN\hspace{1.2ex}MATHEMATIC\kern-1pt
AL\hspace{1.2ex}PHY\kern.4ptSIC\kern-.5ptS}\hfil
     No.\ \Number}\def\@oddfoot{\rm\hfil[\thepage]
     \hfil}\def\@evenhead{}\let\@evenfoot\@oddfoot}
 
\def\cite{\@ifnextchar [{\@tempswatrue\@Rcitex}{\@tempswafalse\@Rcitex[]}}
 
\def\@Rcitex[#1]#2{\if@filesw\immediate\write\@auxout{\string\citation{#2}}\fi
  \def\@citea{}\@cite{\@for\@citeb:=#2\do
    {\@citea\def\@citea{,\penalty\@m\,}\@ifundefined
       {b@\the\value{paPer}R\@citeb}{{\bf ?}\@warning
       {Citation `\@citeb' on page \thepage \space undefined}}%
\hbox{\csname b@\the\value{paPer}R\@citeb\endcsname}}}{#1}}
 
\catcode`\@=12
 
\newcounter{paPer}
\let\Rlabel=\label
\let\Rbibitem=\bibitem
\let\Rref=\ref
\let\Rpageref=\pageref
\def\label#1{\expandafter\Rlabel{\the\value{paPer}R#1}}
\def\bibitem#1{\expandafter\Rbibitem{\the\value{paPer}R#1}}
\def\ref#1{\expandafter\Rref{\the\value{paPer}R#1}}
\def\pageref#1{\expandafter\Rpageref{\the\value{paPer}R#1}}

\def\eMM{\hbox{}}

\def\yyMM{\rule{0ex}{4ex}}

\def\YYMM{\rule{0ex}{6em}}

\newtoks\TITsi
\newtoks\TITsii
 
\def\title#1{\def\TITs{\normalsize\bf\YYMM #1\\[1.75ex]}\def\RHD{#1}}

\def\author#1{\autMM{#1}\def\LHD{#1}}
\def\and{{\rm\lowercase{and}}}
 
\def\autMM#1{\TITsii={\normalsize\sc\yyMM #1\\}%
\TITsi=\expandafter{\TITs}\edef\TITs{\the\TITsi\the\TITsii}}
 
\def\address#1{\TITsii={\yyMM\small\rm #1\\}\TITsi=\expandafter{\TITs}%
\edef\TITs{\the\TITsi\the\TITsii}}
 
\def\received#1{\TITsii={\yyMM\small\rm({\it Received #1\/})}%
\TITsi=\expandafter{\TITs}\edef\TITs{\the\TITsi\the\TITsii}}
 
\def\headtitle#1{\def\RHD{#1}}
\def\headauthor#1{\def\LHD{#1}}
\def\listas#1#2{\addcontentsline{con}{section}{{\sc #1: }{\rm #2}}}

\def\abstract#1{\begin{center} \TITs \end{center}
       \vspace{.1ex}
       \eMM\hfill\parbox[t]{118mm}{\small\hspace*{3ex}%
       #1 \rule[-1.1em]{0em}{1.1em}\par}\par
       \markright{\RHD}
       \markboth{\LHD}{\RHD}}
 
\def\startpaper{%
       \cleardoublepage
       \setcounter{section}{0}
       \stepcounter{paPer}
       \setcounter{equation}{0}
       \setcounter{footnote}{0}
       \setcounter{figure}{0}
       \setcounter{table}{0}
       \def\theequation{\arabic{equation}}
       \def\thefootnote{\arabic{footnote}}
       \thispagestyle{romp}}
 
\def\Year{1996}
\def\Volume{38}
\def\Number{3}
 
\def\PROCcont{\cleardoublepage\thispagestyle{empty}
       \markright{}\markboth{}{}
       \normalsize\rm
       \vspace*{-3em}
       \addtolength{\baselineskip}{-0.35pt}
       \hspace*{\fill}{\large\rm
         Contents of the Volume \Volume, Number \Number}\hspace*{\fill}
       \par\vspace{.5em}
       \par\noindent}
 
\def\endpaper{\relax}

\def\BaddIS{\addtolength{\itemsep}{-1.9ex}}

\font\bBB=msbm10

\def\bBC{\mbox{\bBB C}}

\def\bBR{\mbox{\bBB R}}

\def\bBZ{\mbox{\bBB Z}}

\def\1{{\mathchoice{\rm 1\mskip-4mu l}{\rm 1\mskip-4mu l}%
{\rm 1\mskip-4.5mu l}{\rm 1\mskip-5mu l}}}
\mathchardef\Gamma="7100
\mathchardef\Delta="7101
\mathchardef\Theta="7102
\mathchardef\Lambda="7103
\mathchardef\Xi="7104
\mathchardef\Pi="7105
\mathchardef\Sigma="7106
\mathchardef\Upsilon="7107
\mathchardef\Phi="7108
\mathchardef\Psi="7109
\mathchardef\Omega="710A

\begin{document}


\startpaper
 
\title{MAC~LANE METHOD IN THE INVESTIGATION OF MAGNETIC TRANSLATION
GROUPS}
\headtitle{MAC~LANE METHOD IN THE INVESTIGATION OF MTG's}
\author{Wojciech Florek}
\headauthor{W.~FLOREK}
\listas{W.~Florek}{Mac~Lane Method in the Investigation of Magnetic
  Translation Groups}
\address{Institute of Physics, A. Mickiewicz University,
ul. Matejki 48/49, 60--769 Pozna\'n, Poland\\
e-mail: {\tt florek@plpuam11.amu.edu.pl}
}
\received{February 23, 1996}
 
\abstract{
 Central extensions of the three-dimensional translation group
$T\simeq\bBZ^3$ by the unitary group $U(1)$ (a group of factors) are
considered within the frame of the Mac~Lane method. All nonzero vectors
${\bf t}\in T$ are considered to be generators of $T$. This choice leads
to very illustrative relations between the Mac~Lane method and Zak's
approach to magnetic translation groups. It is shown that factor systems
introduced by Zak and Brown can be realized only for the unitary group
$U(1)$ and for some of its finite subgroups.
}
 
\section{Introduction}
 Much attention is paid recently to two-dimensional systems due their
relations with high-$T_c$ superconductors, anyons, the Hall effect etc.
(see e.g.\ \cite{aoki}).
The behaviour of electrons in periodic potentials and a constant magnetic
field is one of the most important problems. Brown and Zak showed
independently [1--3],
that the magnetic field has to be aligned along a vector of the
crystal lattice and the translations in the plane perpendicular to this
vector do not commute.  Hence, it is enough to consider two-dimensional
translation groups (lattices) to determine properties of the so-called {\em
magnetic translation groups}\/ ({\bf MTG}'s). However, this requirement
(${\bf H}\parallel{\bf t}$) follows from the periodic boundary conditions
imposed on the three-dimensional infinite lattice. 
In this paper, the
infinite (three-dimensional) lattice is considered.
 
It can be shown that the {\bf MTG} is a central extension of the translation
group $T\simeq\bBZ^3$ by a group of factors $G$ (see, e.g.,
\cite{florek1,florek2}); here the case $G=U(1)$ is considered but finite
(cyclic) groups can be also taken into account. Determination of all
nonequivalent (abelian) extensions for given groups, $T$ and $G$ in the
considered case, is equivalent to the determination of the second cohomology
group $H^2(T,G)$, which can be done be means of the Mac~Lane method
[6--8]\@. The form of factor systems $m\colon T\times T\to \bBZ$ obtained by
this method depends strongly on the assumed set of generators of $T$.
In this
paper, the maximal set of generators $A=T\setminus \{{\bf0}\}$ is considered
and relations with Zak's definition of {\bf MTG} are stressed.
 
\section{Choice of Generators}
 Zak \cite{zak1} introduced the {\bf MTG}'s as pairs consisting of a
vector ${\bf t}\in T$ and a path ${\cal P}$ joining the point $O$ 
(the crystal lattice origin) with the point defined by the vector 
${\bf t}$; this path is constructed from $n$ vectors,
 ${\cal P}=({\bf t}_1,{\bf t}_2,\ldots,{\bf t}_n)$ such that
$\sum_{j=1}^n{\bf t}_j={\bf t}$. With each pair $({\bf t}\mid{\cal P})$ the
following operator (in the space of functions
 $\psi\colon\bBR^3\to\bBC$) is associated
 \begin{equation}\label{tau}
 \tau({\bf t}\mid{\cal P})\;=\;
  \exp[-(\mbox{i}/\hbar)\,{\bf t}\cdot({\bf p}-(e/c){\bf A})]
  \exp[-(\mbox{i}/c\hbar)\Phi({\bf t}_1,{\bf t}_2,\ldots,{\bf t}_n)]\,,
 \end{equation}
 where ${\bf A}$ is the vector potential of the magnetic field and
$\Phi({\bf t}_1,{\bf t}_2,\ldots,{\bf t}_n)$ is the magnetic flux through the
polygon enclosed by the vectors ${\bf t}_1,\ldots,{\bf t}_n,-{\bf t}$ (the
symmetric gauge ${\bf A}=({\bf H}\times{\bf r})/2$ is assumed).
 
Such a definition is closely related to the notion of free groups: paths
can be interpreted as words written in an alphabet $X$, which is in a
one-to-one correspondence with the set of nonzero vectors in $T$. Moreover,
the multiplication rule for the operators $\tau$
 \begin{equation}\label{taumul}
 \tau({\bf t}\mid{\cal P})\tau({\bf t}'\mid{\cal P\,}')\;=\;
 \tau({\bf t}+{\bf t}'\mid{\cal P}\bullet{\cal P\,}')\,,
 \end{equation}
 where `$\bullet$' denotes the concatenation of words, i.e.
 $$
 {\cal P}\bullet{\cal P\,}'\;=\;
 ({\bf t}_1,{\bf t}_2,\ldots,{\bf t}_n)\bullet
 ({\bf t}'_1,{\bf t}'_2,\ldots,{\bf t}'_{n'})\;=\;
 ({\bf t}_1,{\bf t}_2,\ldots,{\bf t}_n,
 {\bf t}'_1,{\bf t}'_2,\ldots,{\bf t}'_{n'})\,,
 $$
 is an analogue of the multiplication rule in a free group.
 
On the other hand, the Mac~Lane method consists in replacing an exact
sequence
 $$
\{1\}\longrightarrow U(1) \longrightarrow \mbox{\bf MTG} \longrightarrow T
\longrightarrow \{0\}
$$
 by the following one
 $$
\{1\}\longrightarrow R \longrightarrow F \longrightarrow T
\longrightarrow \{0\}\,,
$$
 where $R$ is the kernel of a homomorphism $M\colon F\to T$ and $F$ is a free
group. Therefore, one may expect that there are closer relations between the
Mac~Lane method and the operators $\tau$ introduced by Zak.
 
The very first, and very important, step in the Mac~Lane method is the choice
of the generators of $T$. In this way, the alphabet $X$ of $F$ is determined
by the condition:
 $$
  M\quad\mbox{\rm restricted to}\quad X\quad \mbox{\rm is a bijection.}
 $$
 There are (infinitely) many sets $A$ being generators of
 $T\simeq\bBZ^3$. If ${\bf a}_j$, $j=1,2,3$, denote the basis vectors of
the crystal lattice then the most natural way is to assume that
$A=\{{\bf a}_1,{\bf a}_2,{\bf a}_3\}$ and $X=\{a_1,a_2,a_3\}$ with
$M(a_j)={\bf a}_j$ for $j=1,2,3$. This choice (the smallest $A$) can be
easily generalized to any dimension of crystal lattice and is very
convenient in further calculations (cf.\ \cite{florek2}). However, all
paths constructed from vectors ${\bf a}_j$ consist of sections, which lie
along edges of a crystal lattice, whereas in Zak's approach segments lie
along {\em any}\/ vectors of a lattice. So one may decide to consider $A$
being the set of all nonzero vectors, i.e.
 \begin{equation}\label{Achoice}
   A\;:=\;T\setminus \{{\bf0}\}\,,
 \end{equation}
 where ${\bf0}$ denotes the zero vector in $T$. A letter of the alphabet $X$
corresponding to a given vector ${\bf t}\in T$ will be denoted by $t$, so
$M(t)={\bf t}$. Since $A$ contains {\em all}\/ nonzero vectors, $X$
also contains the letter $(-t)$ such that $M(-t)=-{\bf t}$. On the other
hand, $F$ being a~free group generated by the alphabet $X$ contains the
inversions of all letters $t\in X$, i.e.\ elements $\bar{t}$ such that
 $$
 \bar{t}t\;=\;t\bar{t}\;=\;1_F,\qquad \forall\,t\in X\,.
 $$
 Of course, in the group $F$ the symbols $\bar{t}$ and $(-t)$ denote
 {\em different}\/ elements, but they belong to the same coset
(the same counter-image) determined by $R=\mbox{\rm Ker}\,M$, since
 $$
 M(\bar{t})\;=\;M(-t)\;=\;-{\bf t}\,.
 $$
 Hence, the words $(-t)t$ belong to $F$, in spite of the fact that they
correspond to trivial loop $(-{\bf t},{\bf t})$ --- the area enclosed by
such a loop is simply~0. To avoid these loops one may consider a
slightly smaller set of generators containing only one vector from each pair
$\{{\bf t},-{\bf t}\}$, ${\bf t}\neq{\bf0}$. If a vector ${\bf t}$ is
represented by $t\in X$ then $-{\bf t}$ is represented by
$\bar{t}\in\overline{X}$ and the words $(-t)t$ are excluded. However, it
occurs that in further calculations sets $X$ and $\overline{X}$ have to be
considered separately and in each case all four possible products, $tt'$,
$\bar{t}t'$, $t\bar{t'}$, and $\bar{t}\,\bar{t'}$ must be taken into
account. This leads to a rather cumbersome calculations and, therefore, the
set
$A$ of the translation group $T$ generators is chosen to be
$T\setminus\{{\bf0}\}$\@. Those inconvenient loops, corresponding to two
antiparallel vectors ${\bf t}$ and $-{\bf t}$, will be excluded in another
way.
 
\section{Factor Systems}
 In the next step of the Mac~Lane method one chooses representatives of
 {\em right}\/ cosets $Rf$. It is important to do it in such a way that
these represenatives form the so-called Schreier set
\cite{florek1,maclane1,lulek,kurosh}. In other words, one defines a
mapping $\psi\colon T\to F$ such that $M\circ\psi=\mbox{id}_T$. In the
case considered the simplest solution is to put
 \begin{equation}\label{Sset}
  \psi({\bf t})\;=\;t\qquad\mbox{\rm and}\qquad \psi({\bf0})\;=\;1_F\,.
 \end{equation}
 The mapping $\psi$ determines also the so-called {\em choice function}\/
 \begin{equation}\label{beta}
  \beta\;=\;\psi\circ M\,,
 \end{equation}
  where $M$ is the above introduced homomorphism, which calculates the value
of a given word $f\in F$ in the group $T$. Of course, words in $F$ are in
one-to-one correspondence with the paths ${\cal P}$ introduced by Zak,
whereas elements of $R=\{f\mid M(r)={\bf0}\}$ correspond to loops 
${\cal L}$.
 
To determine an alphabet $Y$ of the kernel $R$ one has to consider all
products
 \begin{equation}\label{rhofac}
 \psi({\bf t})\psi({\bf t}')\overline{\psi({\bf t}+{\bf t}')}
  \;=\;\varrho({\bf t},{\bf t}')\,,
 \end{equation}
 where the last symbol denotes a factor belonging to $R$. Except for
general case
 \begin{equation}\label{triangle}
  \varrho({\bf t},{\bf t}')\;=\;tt'\overline{\beta(tt')}\,,
 \end{equation}
 there are two special cases:
 \begin{itemize}
 \item ${\bf t}$ or ${\bf t}'$ is equal to ${\bf0}$, so that
 \begin{equation}\label{normal}
    \varrho({\bf t},{\bf0})\;=\;\varrho({\bf0},{\bf t})\;=\;1_F,\,
 \end{equation}
  which means that the factor system $\varrho$ is normalized;
 \item ${\bf t}'\;=\;-{\bf t}$, so that
 \begin{equation}\label{biangle}
    \rho({\bf t},-{\bf t})\;=\;t(-t)\,.
 \end{equation}
 \end{itemize}
The words determined by the relations (\ref{triangle}) and (\ref{biangle})
form
the alphabet $Y$ of the kernel $R$, i.e.\ each word in $R$ can be written as
a (finite) product of such elements. It is easy to notice that these letters
are determined by two nonzero vectors ${\bf t}$ and ${\bf t}'$, which define
the following triangle:
 \begin{center}
 \setlength{\unitlength}{1mm}
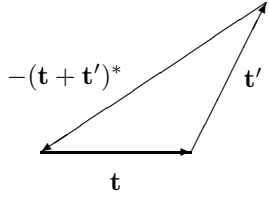
 \begin{picture}(40,25)
 \put(0,5){\vector(1,0){20}}
 \put(10,0){\makebox(0,0)[b]{${\bf t}$}}
 \put(20,5){\vector(1,2){10}}
 \put(27,15){\makebox(0,0)[l]{${\bf t}'$}}
 \put(30,25){\vector(-3,-2){30}}
 \put(11,15){\makebox(0,0)[r]{$-({\bf t}+{\bf t}')^*$}}
\end{picture}
\end{center}
 where an asterisk denotes that a vector $-{\bf t}$ is the image (under the
homomorphism $M$) of $\bar{t}$ and not of $(-t)$\@. In a similar way, ${\bf
t}^*$ denotes the image of $\overline{(-t)}$ to distinguish it from the
image of $t$\@.  Of course, for ${\bf t}'=-{\bf t}$ this triangle is
trivial. Each triangle ${\cal T}({\bf t},{\bf t}')$ encloses an area $({\bf
t}\times{\bf t}')/2$ and each loop in $T$ can be constructed from such
triangles (in many ways) by adding pairs $({\bf t},-{\bf t}^*)$.
 
Factor systems $m\colon T\times T\to U(1)$ are images of $\varrho$ under the
so-called operator homomorphisms $\phi\colon R\to U(1)$, which satisfy the
condition\footnote{To determine all possible operator homomorphisms $\phi$
it is enough to consider $f=t\in X$.}
 \begin{equation}\label{ophom}
  \phi(y)\;=\;\phi(fy\bar{f}),\qquad \forall\: y\in Y,\,f\in F\,.
 \end{equation}
 This requirement results in the following restrictions:
 \begin{enumerate}
  \item $\phi(t(-t))\;=\;\phi((-t)t)$ for any $t\in X$;
  \item $\phi(t(-t))$ is determined by the values of $\phi$ for two
  triangles:
  $$
  {\cal T}({\bf t}',{\bf t})\quad \mbox{\rm and}\quad
   {\cal T}({\bf t}'+{\bf t},-{\bf t})
  $$
  for {\em any}\/ ${\bf t}'\in A$, namely
 $$
   \phi(t(-t))\;=\;\phi(t't\overline{\beta(t't)})
      \phi(\beta(t't)(-t)\bar{t'})\,.
 $$
 \item $\phi(tt'\overline{\beta(tt')})$ is symmetric under any cyclic
  permutation of the arguments:
 $$
  \phi(tt'\overline{\beta(tt')})
  \;=\;\phi(t'\overline{\beta(tt')}t)
  \;=\; \phi(\overline{\beta(tt')}tt')\,.
  $$
  \item For any three letters $x,t,t'\in X$ the following condition has to be
satisfied
 $$
  \phi(tt'\overline{\beta(tt')})\;=\;
  \phi(xt\overline{\beta(xt)})\phi(\beta(xt)t'\overline{\beta(xtt')})
  \phi\Bigl[\overline{x\beta(tt')\overline{\beta(xtt')}}\Bigr]\,;
 $$
 If this result is illustrated by vectors of the crystal lattice, it is easy
to notice that this equation corresponds to the relation between areas,
considered as vectors, of a tetrahedron walls.
 \end{enumerate}
 
Let $m_\phi:=\phi\circ\rho$, i.e.
 \begin{equation}\label{mfac}
  m_{\phi}({\bf t},{\bf t}')\;=\;\phi(\varrho({\bf t},{\bf t}'))
 \end{equation}
 be a factor system determined by the operator homomorphism $\phi$.  If
$\gamma\colon F\to U(1)$ is a homomorphism (i.e.\ it is determined by
the values of
$\gamma(t)$, $t\in X$) then the operator homomorphisms $\phi$ and
$\phi\gamma$,
$(\phi\gamma)(r)=\phi(r)\gamma(r)$, determine equivalent factor systems
\cite{maclane1,lulek}. Choosing $\gamma$ one can obtain factor systems in the
most convenient form. If $\phi$ is an operator homomorphism with
$\phi(t(-t))=\exp({\rm i}\alpha_t)$ then taking $\gamma$ determined by
$\gamma(t)=\gamma((-t))=\exp(-{\rm i}\alpha_t/2)$ one obtains a new operator
homomorphism\footnote{Note that for a finite subgroups of $U(1)$ with an even
number of elements it may be impossible to determine such a homomorphism
$\gamma$.} (a new factor system) with $(\phi\gamma)(t(-t))=1$, i.e.
 $$
 m_{\phi\gamma}({\bf t},-{\bf t})
  \;=\;m_{\phi\gamma}(-{\bf t},{\bf t})\;=\;1\,.
 $$
 In this way, the ``nonphysical'' loops $t(-t)$ are ``removed'' from
consideration and for such factor systems one may always add a~letter
$t(-t)\in Y\subset R$ to any element $r\in R$, since
 $$
 (\phi\gamma)(rt(-t))\;=\;(\phi\gamma)(r)(\phi\gamma)(t(-t))
  \;=\;(\phi\gamma)(r)\,.
 $$
 In particular,
 \begin{equation}\label{change}
 \phi(tt'\overline{\beta(tt')})
  \;=\;\phi(tt'\overline{\beta(tt')}\beta(tt')(-t''))
  \;=\;\phi(tt'(-t''))\,,
 \end{equation}
 where $(-t'')=\psi(-{\bf t}'')$ and ${\bf t}''={\bf t}+{\bf t}'=M(tt')$. It
means that depicting elements of $R$ by loops in the crystal lattice one may
omit an asterisk if: (i) only values of an operator homomorphism are
relevant and (ii) a letter $\bar{t}\in \overline{X}$ occurs at the beginning
or at the end of a word $r\in R$. The first condition is obviously satisfied
here, where the second follows from the fact that each $r\in R$ can be
written using the letters $t(-t)$ (with $\phi(t(-t))=1$) and
$(tt'\overline{\beta(tt')})$ (moreover, one may permute arguments of $\phi$
in the latter case).
 
Further analysis is rather tedious and consists of considering
colinear (parallel), coplanar and general (non-coplanar) triples of
lattice vectors ${\bf t}_1, {\bf t}_2, {\bf t}_3$ in order to show that for
each operator homomorphism there exists an equivalent homomorphism given by
 \begin{equation}\label{factm}
 \phi(tt'\overline{\beta(tt')})
  \;=\;\exp\{-{\rm i}({\bf t}\times{\bf t}')\cdot{\bf h}/2\}\,,
 \end{equation}
 where $\times$ denotes the vector product
 $$
    ({\bf t}\times {\bf t}')_j=\varepsilon_{jkl}t_kt'_l\,,
 $$
 and ${\bf h}\in\bBR^3$ is any three-dimensional vector. Please note that
operator homomorphisms (so factor systems, too) determined by different ${\bf
h}\in\bBR^3$ may be equivalent (see discussion in the next section). The
factor $1/2$ was introduced to stress that it is possible to write each
factor system (each operator homomorphism) in such a form if and only if the
equation $\alpha=2\beta$ can be solved (in other words, the equation ${\rm
e}^{{\rm i}\alpha}={\rm e}^{2{\rm i}\beta}$). Of course, it is possible in
$U(1)$ but may be impossible in its (finite) subgroups. Comparing this with
the formula (9) in \cite{brown} (or (14) in \cite{zak1}) which reads
 $$
 T({\bf t})T({\bf t}')=\exp[(-{\rm i}/2)({\bf t}\times{\bf t}')\cdot
  (e{\bf H}/\hbar c)]\,,
 $$
 one can see that the parameter ${\bf h}$ should be interpreted as
 \begin{equation}\label{hH}
 {\bf h}=\frac{e{\bf H}}{\hbar c}\,.
 \end{equation}
 
\section{Final Remarks}
It has been shown that the magnetic translation group introduced by Zak
\cite{zak1} is in fact a free group $F$ generated by nonzero vectors of the
(ordinary) translation group $T\equiv\bBZ^3$, namely a path ${\cal P}$
associated with a vector ${\bf t}\in T$ is a word written in this free group.
The homomorphism $M\colon F\to T$ ``calculates'' the value of this word in
$T$,
giving the vector ${\bf t}$. One-letter words $t=\psi({\bf t})$ can be chosen
to form the alphabet of $F$ and, together with the unit element
$1_F=\psi({\bf0})$, form the set of (right) coset representatives. This group
is too big for physics applications since factor systems (labelled by ${\bf
h}$) depend on the area of a polygon enclosed by vectors ${\bf t}_1$, ${\bf
t}_2$, \ldots, ${\bf t}_n$, $-{\bf t}$, belonging to the path ${\cal P}$, 
in the plane perpendicular to ${\bf h}$.  On the other hand, the central
extension of $T$ by $U(1)$ introduced in this paper can be also too big in
some cases. Let $\langle\exp({\rm i}\alpha),{\bf t}\rangle$ denotes an
element of this extension. The multiplication rule reads
 $$
 \langle\exp({\rm i}\alpha),{\bf t}\rangle
 \langle\exp({\rm i}\alpha'),{\bf t}'\rangle\;=\;
 \langle\exp\{{\rm i}[\alpha+\alpha'-({\bf t}\times{\bf t}')\cdot{\bf h}]\}
 ,{\bf t}+{\bf t}'\rangle\,.
 $$
 Since
 $$
 {\bf t}\times{\bf t}'\;=\;
  \frac{V}{2\pi}(k_1{\bf a}_1^*+k_2{\bf a}_2^*+k_3{\bf a}_3^*)\,,
 $$
 where ${\bf a}_j^*$ are vectors of the reciprocal lattice, $V$ is the
volume of the unit cell, and all the $k_j$ are integers, then for
 $$
  {\bf h}\;=\;\frac{4\pi}{V}(q_1{\bf a}_1+q_2{\bf a}_2+q_3{\bf a}_3)\,,
 $$
 where $q_j$ are rational numbers, one obtains
 \begin{equation}\label{rational}
 {1\over2}({\bf t}\times{\bf t}')\cdot{\bf h}\;=\;
  2\pi(k_1q_1+k_2q_2+k_3q_3)\,.
 \end{equation}
 Therefore, the factors take only a finite number of different values and,
moreover, these values form a group $G\subset U(1)$ generated by three
complex numbers
 $$
  \exp(2\pi{\rm i}q_j),\qquad j=1,2,3.
 $$
 In such a case it is enough to consider an extension of $T$ by this group.
Moreover, the formula (\ref{rational}) shows that factor systems are periodic
with respect to the parameters~$q_j$ (even if they are not rational) with
identical periods equal to 1, i.e.\ the periods of ${\bf h}$ are equal to
$(4\pi/V)|{\bf a}_j|$. Substituting it into (\ref{hH}) one finds that the
physical properties of the considered system are periodic with respect to the
magnetic {\bf H} field with periods
 $$
 H_j=\frac{2}{V}\cdot\frac{hc}{e}|{\bf a}_j|\,.
 $$
 which agrees with the results of Brown \cite{brown} and Zak \cite{zak2}. The
quantity $hc/e$ is the elementary quantum of magnetic flux (fluxon), and the
above formulae show that a finite group $G'$ may be considered if all the
$q_j$ are rational numbers, which means that the magnetic flux through areas
enclosed by any vectors of the Bravais lattice is equal to a rational
multiplicity of the fluxon.
 
It is interesting to consider a factor system, which is equivalent to the
previous one, given by (\ref{factm})
 $$
  m'({\bf t},{\bf t'})\;=\;
  \exp\{-{\rm i}({\bf t}\wedge{\bf t}')\cdot{\bf h}\}\,,
 $$
 where
 $$
  {\bf t}\wedge{\bf t}'\;=\;t_2t'_3{\bf a}_1^*+t_3t'_1{\bf a}_2^*
  +t_1t'_2{\bf a}_3^*\,.
 $$
 These systems are equivalent since they differ by trivial factor system
 $$
 \theta({\bf t},{\bf t'})
  \;=\;\exp\{-{\rm i}({\bf t}\vee{\bf t}')\cdot{\bf h}/2\}\,,
 $$
 where
 $$
  {\bf t}\vee{\bf t}'\;=\;(t_2t'_3+t_3t'_2){\bf a}_1^*
   +(t_3t'_1+t_1t'_3){\bf a}_2^*+(t_1t'_2+t_2t'_1){\bf a}_3^*\,
 $$
 which is generated by a mapping $\eta\colon T\to U(1)$
 $$
 \eta({\bf t})\;:=\;\exp\{{\rm i}({\bf t}\wedge{\bf t})\cdot{\bf h}/2\}
 $$
 according to the well-known formula
 $$
 \theta({\bf t},{\bf t'})\;=\;
  \eta({\bf t}) \eta({\bf t}')/\eta({\bf t}+{\bf t}')\,.
 $$
 Three important facts have to be stressed:
 
 (i)
  This equivalence of factor systems can be established only for the group
$U(1)$ and its subgroups, in which the equation ${\rm e}^{{\rm i}\alpha}={\rm
e}^{2{\rm i}\beta}$ can be solved. The factor system $m'$ is more general,
since this condition need not be satisfied. (This factor system is a
general one, in a sense: it can be introduced for any $G\subset U(1)$,
whereas the second one only for groups fulfilling the above condition.)
 
  (ii)
  It leads to nontrivial factors for pairs $({\bf t},-{\bf t})$
(words $t(-t)$) since
  $$
  m'({\bf t},-{\bf t})
  \;=\;\exp\{{\rm i}({\bf t}\wedge(-{\bf t}))\cdot{\bf h}/2\}
  \;=\;\exp\{-{\rm i}(t_2t_3{\bf a}_1^*+t_3t_1{\bf a}_2^*+t_1t_2{\bf a}_3^*)
  \cdot{\bf h}\}.
  $$
 
 (iii)
 However, in both cases one obtains
 $$
 \langle 1,{\bf t}\rangle \langle 1,{\bf t}'\rangle\;=\;
 \langle 1,{\bf t}'\rangle \langle 1,{\bf t}\rangle
  \exp\{-{\rm i}({\bf t}\times{\bf t}')\cdot{\bf h}\}\,.
 $$
 This means that these factor systems correspond to two different
decompositions of the above commutator, expressed by the vector product
$\times$, into two parts. The first decomposition is (anti)symmetric,
whereas the second is asymmetric and resembles the Landau gauge. Recalling
Brown's work \cite{brown}, in which the ray (projective) representation
of $T$ was introduced, one can see why his approach is appropriate only in
the case of the symmetric gauge ${\bf A}=({\bf H}\times{\bf r})/2$ (a more
general condition is given by formula (12) in \cite{zak1}): Two ray
representations with different factor systems are nonequivalent and one has
to define another ray representation of $T$ to cover other gauges (other
vector potentials~${\bf A}$).
 
Finally, it should be emphasised that all the above considerations can be
applied to a crystal lattice of any dimension (which was partially done in
\cite{florek2}) and with application of tensor algebra introducing in a
proper way co- and contravariant tensors, contractions, tensor products,
polyvectors (multivectors) etc. As was mentioned at the beginning, one may
limit oneself to two-dimensional lattices (with the magnetic field
perpendicular to the crystal plain, so it can be considered as a scalar),
but it is interesting and important to learn more about the algebraic
structure
of magnetic translation groups for the sake of a proper understanding of
the physical consequences.
 
\section*{Acknowledgements}
 I would like to thank S.~Wa{\l}cerz for fruitful discussions. I am also
grateful to the organizers of the 28th Symposium on Mathematical Physics in
Toru\'n for their hospitality.
 
This work was supported by the Polish State Committee for Scientific Research
(KBN) within the project No PB 201/P3/94.

\endpaper
\end{document}